\documentstyle[prb,eqsecnum,aps,epsf]{revtex}
\font\bba=msbm10 scaled 1200
\font\bbb=msbm8 
\newfam\bbfam
\textfont\bbfam=\bba
\scriptfont\bbfam=\bbb
\def\bb{\fam\bbfam\bba}
\def\R{{\bb R}}

\begin{document}
\draft
\title
{Monte Carlo simulations of the  screening potential of the Yukawa 
one-component plasma.}
\author{J.-M. Caillol \thanks{e-mail : Jean-Michel.Caillol@th.u-psud.fr}}
\address{
	 LPT - CNRS (UMR 8627) \\
	 Bat. 210,
	 Universit\'e de Paris Sud \\
	 F-91405 Orsay Cedex, France}

\author{D. Gilles \thanks{e-mail: Dominique.Gilles@bruyeres.cea.fr}}
\address{Commissariat \`a l'Energie Atomique\\
	  BP12                              \\
	  91680 Bruy\`eres-le-Ch\^atel, France}                  
\date{\today}
\maketitle
\begin{abstract}
A Monte Carlo scheme to sample the screening potential H(r) of Yukawa plasmas 
notably at short distances is presented. This scheme is based on an importance
 sampling
technique. Comparisons with former results  for the Coulombic one-component
 plasma are given. Our Monte Carlo simulations yield an accurate estimate 
 of H(r)
as well for short range and long range interparticle distances.
 
\end{abstract}
 
\section{Introduction}
\label{intro}
In this work we present a Monte Carlo scheme devised to compute the pair 
distribution function $g(r)$ of a strongly coupled plasma, notably at
 short distances,i.e. for values of $r$ smaller or of the order of the ionic
 radius, 
 with a high accuracy. The model considered in our study is the Yukawa 
 One-Component plasma (YOCP), i.e. a system made of $N$ identical classical
  ions of charge $Z e$ immersed in an uniform neutralizing background of 
  electrons. The effective interaction between two ions is supposed to be of 
  the form  
$v_{\alpha}(r)=(Ze)^2 \  y_{\alpha}(r)$ where 
$y_{\alpha}(r)= \exp(-\alpha r)/r$ ($\alpha \ge 0 $) is the Yukawa potential
and $\alpha$  the screening parameter. In the limit $\alpha \to 0$ the YOCP 
reduces to the well-known coulombic one-component plasma (OCP)\cite{Hansen}. 
The configurational potential energy in a domain $\Lambda$ of the ordinary 
space $\R^{3}$ can thus be written as  \cite{Caillol} 
\begin{eqnarray}
\label{confi1}
V_{\Lambda}(1,\ldots,N)&=& 
\frac{(Ze)^{2}}{2}\; \sum_{i \neq j}^{N} y_{\alpha}(r_{ij})  + Ze
\sum_{i=1}^{N}\int_{\Lambda} d^{3}\mbox{\boldmath $r$ }
 \rho_{B} y_{\alpha}(|\mbox{\boldmath $r$}-\mbox{\boldmath $r$}_{i}|)
  \nonumber \\
&+&
\frac{1}{2}
\int_{\Lambda } d^{3}\mbox{\boldmath $r$ }
d^{3}\mbox{\boldmath $r'$} \rho_{B}^{2} y_{\alpha}(|\mbox{\boldmath $r$}-
\mbox{\boldmath $r$}'|) + N (Ze)^{2} {\cal E} \; , 
\end{eqnarray}
where $\rho_{B}= -N Ze/\Lambda$ is the uniform charge density of the 
background.
The constant  
${\cal E}$ which appears in the r.h.s of eq.\ (\ref{confi1})
fixes the zero of energy and reads as  
\begin{equation}
 {\cal E}=\frac{1}{2} \lim_{r \rightarrow 0}\;
[y_{\alpha}(r)-\frac{1}{r}] =-\alpha /2 \; .
\label{cte}
\end{equation}
In the thermodynamic limit the thermodynamic, structural, and  dynamical 
properties of the YOCP depend solely upon  two dimensionless parameters, 
namely
the coupling parameter $\Gamma = \beta (Ze)^{2}/a $ and the reduced screening
parameter 
$\alpha ^{*}= \alpha a $ where $\beta = 1/k_{B}T$ ($k_{B}$ Boltzmann constant,
 $T$ temperature) and $a$ is the ionic radius ($4\pi 
\rho a^{3}/3=1$, where $\rho = N/\Lambda$ is the number density of particles).

The thermodynamic properties of the YOCP are well-known nowadays thanks to 
extensive Monte Carlo (MC) simulations performed either within periodical
\cite{Hamaguchi} or hyperspherical \cite{Caillol}  boundary conditions.
Much less is known about $g(r)$ and the related screening function $H(r)$ 
defined as 
\begin{equation}
\label{H}
g(r) = \exp \left( -\Gamma a  y_{\alpha}\left(r\right) +
 \Gamma H\left(r \right) \right) \; .
\end{equation}
$H(r)$ plays an important role in estimating the enhancement factors for the 
thermonuclear reaction rates\cite{Ichi}.
As for $r \to 0$  $g(r)\sim \exp(-\Gamma a
y_{\alpha}(r))$ the values of $g(r)$ for $r \to 0$ are extremely small for 
large
$\Gamma$'s, which precludes a numerical study by means of standard MC 
simulations and biased MC schemes are therefore unavoidable.
Such a scheme is presented in  next sec\ (\ref{screening}), 
it is a synthesis of two biased schemes applied formerly in the 
determination of the cavity function of hard spheres \cite{Patey} in the
one hand, and in the calculation of the screening function $H(r)$ of the
OCP in the other hand \cite{Ogata}.

Results of MC simulations are reported in sec\ (\ref{MC}) both for the OCP 
and the YOCP cases. They must be considered as preliminary results, our 
ultimate goal being to establish a complete data basis of screening functions
$H(r)$ for a wide range of
$(\Gamma,\alpha^{*})$.
 
\section{Sampling the screening function}
\label{screening}
In the canonical ensemble the pair distribution function $g(r)$ is given by 
\begin{equation}
    \label{gdr}
g(r)= \Lambda \langle
\frac{\int \prod_{i=1}^{N}d\vec{r}_{i}\; 
\delta(\vec{r}-\vec{r}_{12})
\exp\left(-\beta V_{\Lambda}\left(1,2\cdots,N\right)\right)}{\int 
\prod_{i=1}^{N}d\vec{r}_{i}\;\exp(-\beta V_{\Lambda}(1,2\cdots,N))} \rangle 
\; ,
\end{equation}
where the brackets $< \ldots > $ denote a canonical thermal average and
$ \Lambda$ is the volume.
In eq.\ (\ref{gdr}) $\vec{r}_{12}\equiv \vec{r}_{1} - \vec{r}_{2}$ and we 
have implicitly assumed that the system was homogeneous which is verified if
 periodical boundary conditions are adopted. In practice $g(r)$
 can be computed in a standard MC calculation with a good precision only 
for $r>r_{min}$. For instance at $\Gamma \sim 100$ we have
typically $r_{min}/a \sim 1$. In order to compute $g(r)$ for $r<r_{min}$ we 
follow the suggestion of Ogata \cite{Ogata} and rewrite eq.\ (\ref{gdr}) as
\begin{eqnarray}
    \label{gw}
    g(r_{12}) & = & \Lambda \langle
\frac{\int \prod_{i=1}^{N}d\vec{r}_{i}\; 
\delta(\vec{r}-\vec{r}_{12})\exp(-\beta 
V_{\Lambda}(1,2\cdots,N)-\beta w(r_{12}))\exp(\beta w(r_{12}))}{\int 
\prod_{i=1}^{N}d\vec{r}_{i}\;\exp(-\beta V_{\Lambda}(1,2\cdots,N))} \rangle 
\nonumber \\
&\propto &  g_{w}(r_{12}€) \exp(+\beta w(r_{12})) \; .
\end{eqnarray} 
In\ (\ref{gw})  $w(r)$ is {\em a priori} an arbitrary function and Ogata has
 shown  how to take advantage of that to devise an efficient MC biased scheme.  

The function $g_{w}$ supports the following simple physical interpretation.
 Let us consider a mixture made of $(N-2)$ Yukawa charges and two test 
 particles labelled $(1,2)$. These two particles interact with the $(N-2)$
  other ions
 via Yukawa potentials {\em but} their mutual potential energy is defined as 
$w(r_(12))+v_{\alpha}(r_{12})$. For practical purposes, it is clearly clever to choose 
$ \beta w(r)=-\Gamma a y_{\alpha}(r) + \Gamma \overline{H}(r)$ where 
$\overline{H}(r)$ is a good estimate of the {\em true} screening function 
$H(r)$. Indeed it follows from eqs.\ (\ref{H}) and\ (\ref{gw}) that $g_{w}(r) 
\propto \exp \left( \Gamma  \left[ H\left( r \right)- \overline{H}\left(
r \right) \right] \right)$. As a consequence, if $H\sim \overline{H}$, 
then $g_{w}(r)$ is practically constant and therefore easy to determine
 numerically.
However we are only half the way since $g_{w}(r)$ is known only up
to a multiplicative constant. This normalization constant can be reexpressed,
as discussed  by Ogata\cite{Ogata}, as a difference of free energies which 
can be determined in the course of a MC run  as  thermal averages.
However, in the MC simulations of this author,
the variance on these averages turn out to be quite large which precludes 
an accurate estimate. Therefore we adopted the method that Patey and
Torrie devised for computing the cavity function of
hard spheres\cite{Patey}. Suppose that $g_{w}(r)$ has been computed 
by a MC simulation of the mixture described above 
for, let say, $0 < r/a <2$. In practice it can be achieved by the choice
 $ \beta w(r)=-\Gamma a y_{\alpha}(r) + \Gamma \overline{H}(r)$
  for  
$0 < r/a <2$ and $ \beta w(r)=\infty$  for $r/a>2 $. 
Then an {\em unnormalized} pair distribution 
$g_{u}(r)=g_{w}(r) \times \exp( \beta w(r))$
can be computed in the range $0 <r/a < 2$ and compared to  
the {\em normalized} $g_{0}(r)$ obtained by a standard MC simulation. 
This comparison in the range of distances where both $g_{0}(r)$ 
and $g_{u}(r)$ can be determined precisely (i.e. $ r_{min}/a<r/a <2  $)
 yields an accurate determination of the wanted normalization constant.
  Therefore both $g(r)$ and $H(r)$ can be computed accuratly for all $r$'s.
\section{Monte Carlo simulations}
\label{MC}
The scenario of sec.\ \ref{screening} was scrupulously applied in our
MC simulations which were performed within hyperspherical boundary 
conditions\cite{Caillol}. For each pair of $(\Gamma,\alpha^{*})$ a
standard and a biased MC simulation was performed. The runs were 
divided in typically $\sim 20$ sub-runs in order to compute the 
statistical errors by a method of blocks\cite{Frenkel}. In the 
preliminary data reported here the considered systems involved 
$N=500$ particles but a systematic study of finite size effects 
is under way. 

In the case of the OCP we chose for 
$\overline{H}(r)$ the fit of $H(r)$ provided by Ogata\cite{Ogata}.
However a small attractive potential $\delta \overline{H}(r)$ was added to 
$\overline{H}(r)$ at short distances to enhance the sampling 
of $g_{w}(r)$ at small $r$. The form of $\delta \overline{H}(r)$
is of course irrelevant, we retained a quadratic expression for
convenience. In the case of the YOCP the bias function
$\overline{H}(r)$ at $(\Gamma,\alpha^{*}+ \delta \alpha^{*})$ was 
chosen equal to the screening function $H(r)$ obtained for the 
point $(\Gamma,\alpha^{*})$.

\begin{figure}
\begin{center}
\epsfxsize=2.8 truein
\epsfbox{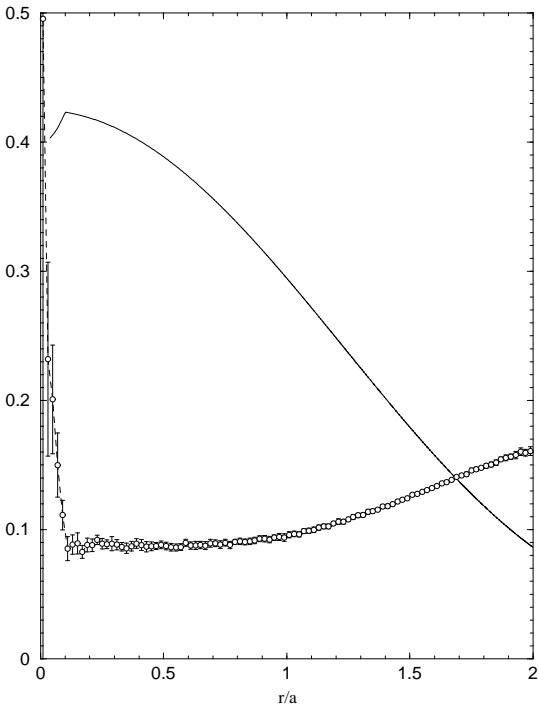}
 \caption{The bias function  $\overline{H}(r)$ (top curve, solid line) used 
at $(\Gamma=40,\alpha^{*}=1)$ to determine the function $g_{w}(r)$
(bottom curve). The error bars on $g_{w}(r)$ correspond to two standard 
deviations. }
\label{fig-gw}
\end{center}
\end{figure}

\begin{figure}
\begin{center}
\epsfxsize=2.8 truein
\epsfbox{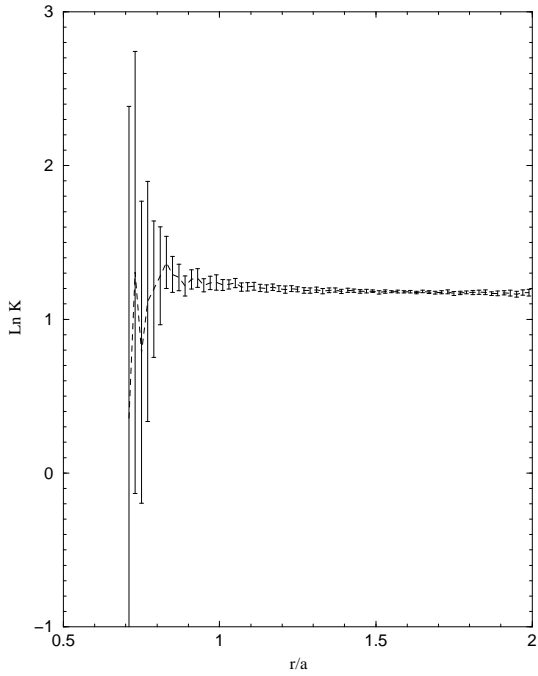}
 \caption{Logarithm of the ratio $K(r)$ of the normalized $g_{0}(r)$ (standard
MC calculation) and the unnormalized $g_{u}(r)$ (biased MC calculation) in the
range $r_{min}<r<2 a$ at $(\Gamma=40,\alpha^{*}=1)$.}
\label{fig-K}
\end{center}
\end{figure}

\begin{figure}
\begin{center}
\epsfxsize=3.2 truein
\epsfbox{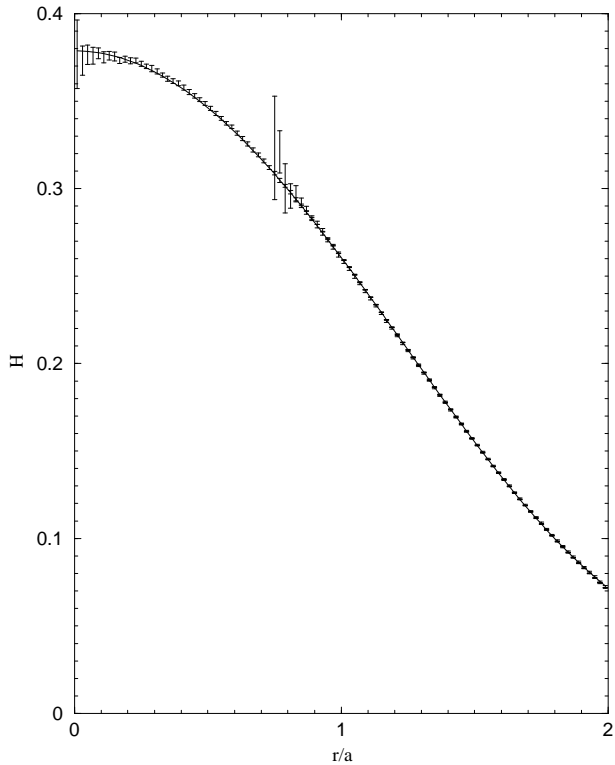}
 \caption{The screening function $H(r)$
at $(\Gamma=40,\alpha^{*}=1)$ for a sample of 
$N=500$ particles. The standard MC calculation allows only the determination of
 $H(r)$ in the
range $r_{min}<r < 2 a$ (curve on the right). The biased MC gives the curve
 in the
range $ r < 2 a$. In the overlapping region the agreement between these two
estimates  is satisfactory. The solid curve is a $6^{th}$ order 
even polynomial of the MC data }
\label{fig-H}
\end{center}
\end{figure}

To illustrate the method, we display in fig.\ (\ref{fig-gw}) 
a plot of the bias function $\overline{H}(r)$ and of the 
unnormalized $g_{w}(r)$ at $(\Gamma=40,\alpha^{*}=1)$. 
As a result of the structure of the bias function $\overline{H}(r)$ at 
short distance
$g_{w}(r)$ exhibits a quite pronounced peak  in this region. 
We show in  fig.\ (\ref{fig-K}) the logarithm of the ratio $K(r)$ of
the unnormalized $g_{u}(r)$ by the properly normalized $g_{0}(r)$ at the same
point $(\Gamma=40,\alpha^{*}=1)$.
As can be seen in the figure the normalization constant of $g_{u}(r)$
can be obtained with a good accuracy by averaging $K(r)$ for $1<r/a<2$.
The resulting function $H(r)$ is displayed in fig.\ (\ref{fig-H}) for $r$ 
varying in the range $0<r/a<2$. Note that in the region
$r_{min}<r/a<2$ where $H(r)$ can be computed by a standard MC 
simulation it coincides almost perfectly with the  $H(r)$
 computed by the biased scheme.
In all cases the screening function can be accurately fitted
 by the simple polynomial 
 $a_{0} +a_{2}(r/a)^2 + a_{4}(r/a)^4 + a_{6}(r/a)^6 $
  which traverses all the error bars. 

\begin{figure}
\begin{center}
\epsfxsize=3.2 truein
\epsfbox{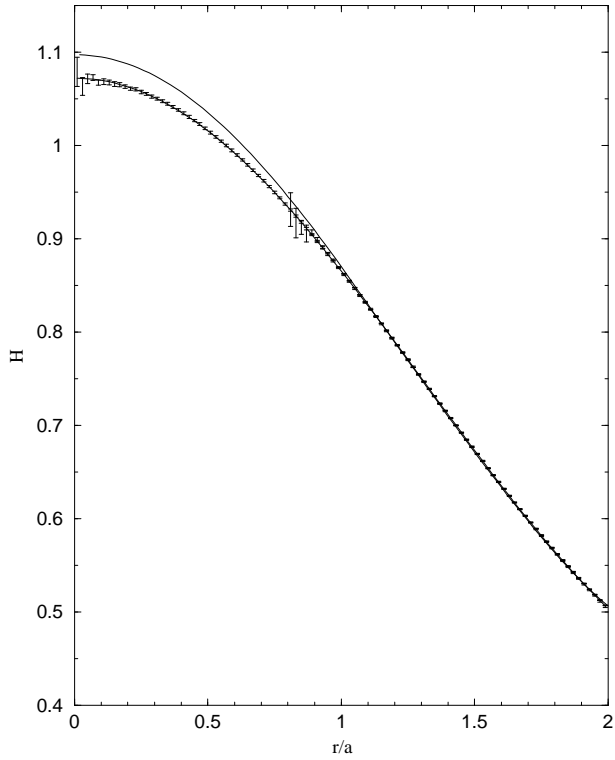}
 \caption{The screening function $H(r)$ for the OCP at $\Gamma=40$. Top curve:
Ogata result. Bottom curves: MC data for a sample of $N=500$ ions (standard and
biased simulations). The solid curve is a $6^{th}$ order 
even polynomial of the MC data. }
\label{fig-OCP-Ogata}
\end{center}
\end{figure}

\begin{figure}
\begin{center}
\epsfxsize=3.2 truein
\epsfbox{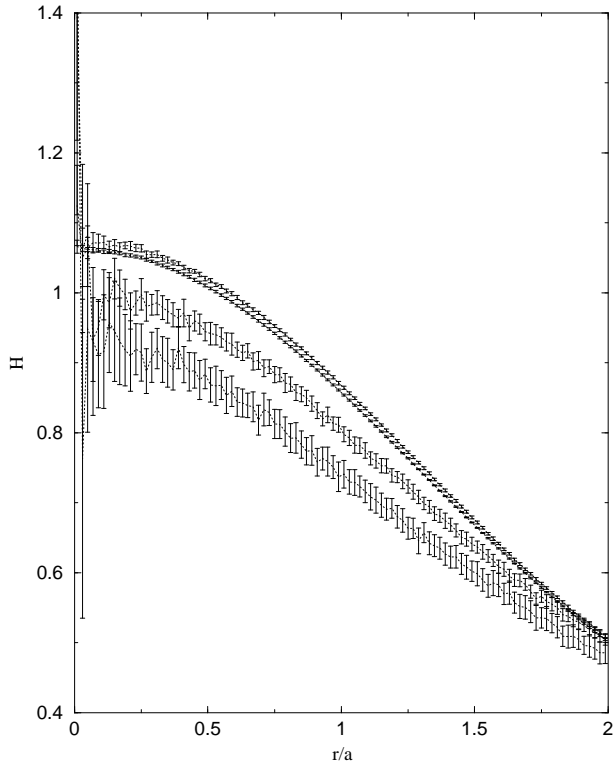}
 \caption{ The screening function $H(r)$ for the OCP. From bottom to top
$\Gamma=1, 2, 10$, and $100$.}
\label{fig-OCP}
\end{center}
\end{figure}

In the case of the OCP, the theoretical value 
$a_{2}=-1/4$\cite{Janco,Rosen,DEWITT} is roughly recovered although 
its numerical value seems to depend strongly upon the number of particles.
For instance at $\Gamma=40$ one finds  $a_{2}=-0.234, -0.239, -0.249$, and
$ -0.25$
for samples of respectively $N=500, 1000, 2000$, and $3000$ particles. 
A more systematic study of these finite volume effects 
will be presented elsewhere.
In the case of the OCP our results for $H(r)$ differ by a small but 
significant amount of those of Ogata (see fig.\ (\ref{fig-OCP-Ogata})), 
but we have no firm conclusion concerning these discrepancies. 
A bunch of $H(r)$ for various $\Gamma$ is displayed in  fig.\ (\ref{fig-OCP}).

\begin{figure}
\begin{center}
\epsfxsize=3.2 truein
\epsfbox{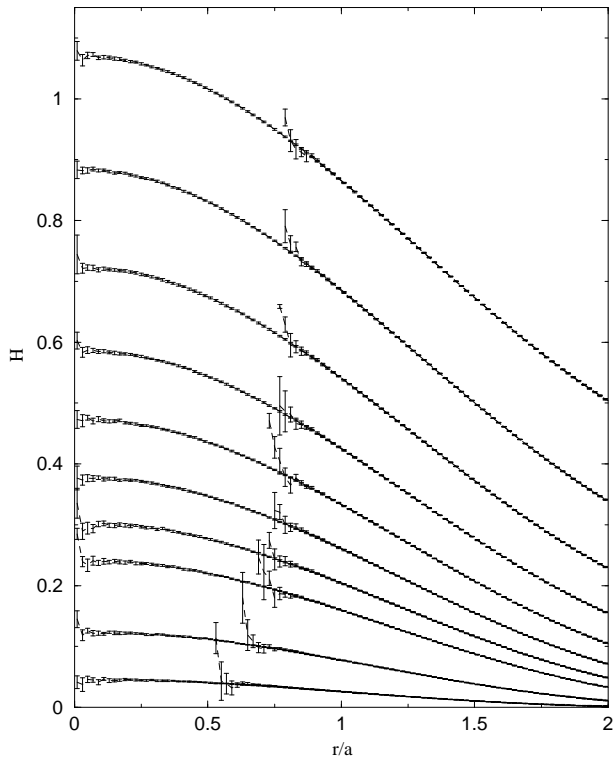}
 \caption{The screening function $H(r)$ for the YOCP at $\Gamma=40$ and some
values of $\alpha^{*}$. From  top to bottom
$\alpha^{*}=0., 0.2, 0.4, 0.6, 0.8, 1., 1.2, 1.4, 2,$ and $ 3$. }
\label{fig-YOCP}
\end{center}
\end{figure}

In the case of the YOCP the amplitude of the function $H(r)$ 
at a given $\Gamma$
decreases steadily as $\alpha$ increases as can be shown 
in fig.\ (\ref{fig-YOCP}) but the shape of the curves at
short distances remains unchanged.  A $6^{th}$ order 
even polynomial perfectly fits the numerical data in the range $0<r/a<2$.
A test of some theoretical predictions by Rosenfeld
and Chabrier\cite{RosenChabrier} concerning $H(r)$ at small $\alpha$ 
is planned for future work.

\end{document}